\documentclass[]{aa}

\usepackage{graphicx}

\usepackage{txfonts}

\usepackage{color}

\usepackage{natbib}

\usepackage[normalem]{ulem}

\bibpunct{(}{)}{;}{a}{}{,} 

\newcommand{\qm}[1]{``#1''}
\newcommand{\fromto}{\,${---}$\,}

\def\sss{\scriptscriptstyle}
\def\U{{\sss \!U}}
\def\L{{\sss \!L}}

\def\nuL{\nu_\L}
\def\nuU{\nu_\U}

\begin{document}
\title{Reversal of the amplitude difference of kHz QPOs in six atoll sources}

\author
{Gabriel T\"or\"ok
}

\offprints{G.~T\"or\"ok,~ terek@volny.cz}

\institute{Institute of Physics, Faculty of Philosophy and Science, Silesian
  University in Opava, Bezru\v{c}ovo n\'{a}m. 13,CZ-74601 Opava, Czech Republic}
  
\date{Received / Accepted}

\abstract
{
Several models have been proposed to explain the twin kilohertz quasi-periodic oscillations (kHz QPOs) detected in neutron-star low-mass X-ray binaries; but when confronting theory with observations, not much attention has been paid so far to the relation between their amplitudes.}
{For six neutron-star atoll sources (namely 4U~1608-52, 4U~1636-53, 4U~0614+09, 4U~1728-34, 4U~1820-30, and 4U~1735-44,) we investigate the relationship between the observed fractional rms amplitudes of the twin kHz QPOs. We discuss whether this relationship displays features that could have a physical meaning in terms of the proposed QPO models.}
{We consider the difference in rms amplitude between the upper and lower kHz QPOs as a function of the frequency ratio $R=\nuU/\nuL$. We compared two data sets. Set I is a collection taken from published data. Set II has values for the rms amplitudes obtained by automatic fitting of all RXTE-PCA observations available up to the end of 
2004, corresponding to continuous segments of observation.}
{\emph{For each of the six sources, we find that there is a point in the R domain around which the amplitudes of the two twin kilohertz QPOs are the same.} We find such a point located inside a narrow interval $R=1.5\pm\,3\%$. Further investigation is needed in the case of 4U~1820-30 and 4U~1735-44 to explore this finding, since we have not determined this point in Set II. There is evidence of a similar point close to R = 1.33 or R = 1.25 in the four sources \mbox{4U~1820-30}, \mbox{4U~1735-44}, \mbox{4U~1608-52}, and \mbox{4U~1636-53}. We suggest that some of these special points may correspond to the documented clustering of the twin kHz QPO frequency ratios.}
{For the sources studied, the rms amplitudes of the two twin peaks become equal when the frequencies of the oscillations pass through a certain ratio $R$, which is roughly the same for each of the sources. In terms of the orbital QPO models (of both the hot-spot and disc-oscillation types), with some assumptions concerning the QPO modulation mechanism, this finding implies the existence of a specific orbit at a particular common value of the 
dimensionless radius, at which the oscillations corresponding to the two twin peaks come into balance. In a more general context, the amplitude difference behaviour suggests a possible energy interchange between the upper and lower QPO modes.}

\authorrunning{G. T\"or\"ok}
\titlerunning{Reversal of the amplitude difference of kHz QPOs in six atoll sources}

\maketitle

%
\section{Introduction}\label{section:introduction}

A number of black-hole and neutron-star sources in low-mass X-ray binaries (LMXBs) show quasi-periodic oscillations (QPOs) in their observed X-ray fluxes, i.e., narrow features (peaks) in their power density spectra (hereafter PDS). The frequencies of these QPOs range from \mbox{$\sim\!10^{-2}$\,Hz} to \mbox{$\sim\!10^{3}$\,Hz}. Here we restrict our attention to the so-called \emph{kHz} (or high-frequency; HF) QPOs, which have frequencies in the range $200\fromto1300\,\mathrm{Hz}$, i.e., of the same order as the frequencies of orbital motion close to the compact object \citep[see][~for a review]{Kli:2006:CompStelX-Ray:}.

In black-hole systems HF QPO peaks are typically detected at constant frequencies characteristic of a given source. When two or more HF QPO frequencies are detected, they usually come in small-number ratios, typically in a \mbox{$3\!:\!2$} ratio \citep[][]{abr-klu:2001, mc-rem:2003}. For neutron-star sources, on the other hand, kHz QPOs often arise as two simultaneously observed\footnote{The term \qm{simultaneously observed} is here applied to continuous observations (or their subsegments) displaying both modes. The integration time needed for these detections ranges from a few seconds to a few dozen minutes.} peaks with frequencies that change over time. This paper focuses on these kHz PDS features, referred to as {twin kHz QPOs}.

From here on, we adopt the convention of calling the two peaks forming twin kHz QPOs the \emph{lower and upper QPO} and denote their frequencies as $\nuL\!<\!\nuU$. The twin kHz QPOs span a wide frequency range and follow a 
nearly linear $\nuU$--$\nuL$ relation for each source~\citep[][]{Bel-Men-Hom:2005:ASTRA:,Abr-etal:2005:RAGtime6and7:CrossRef,Bur:2006:PhD:}.

In this paper we use the twin kHz QPO data obtained with the Proportional Counting Array \citep[PCA;][]{jah-etal:1996} on board the Rossi X-ray Timing Explorer \citep[RXTE;][]{bra-etal:1993}.

\subsection{Orbital models of kHz QPOs and frequency ratio}

Several models have been proposed to explain kHz QPOs, and most of them involve orbital motion in the inner regions of the accretion disc \citep[see][for a recent review]{Kli:2006:CompStelX-Ray:,Lam-Bou:2007:ASSL:ShrtPerBS}.

Among others, two frequently discussed models are based on strong-gravity properties. \citet{Ste-Vie:1999} introduced the ``relativistic precession model'' (which we will refer to as the \emph{RP model}) in which the QPOs are associated with the motion of blobs of matter in the inner parts of the accretion disc. The upper and lower QPO frequencies $\nuU,~\nuL$ are identified with the Keplerian frequency and the relativistic periastron precession frequency, respectively. Slightly after this, \citet{Klu-Abr:2000:PHYRL:} suggested a model in which the twin kHz QPOs arise from non-linear resonance between two modes of \emph{accretion disc 
oscillation}. We will refer to this as the \emph{resonance model}. In the basic version of the resonance model, the upper and lower QPO frequencies, when in a resonant ratio, coincide with the resonant frequencies given by specific combinations of epicyclic frequencies of geodesic test particle orbital motion \citep[see][]{abr-klu:2001,abr-etal:2003,reb:2004,Hor-Kar:2006:ASTRA:TOQPOIntRes,abr-etal:2006,swe:2008}.

Frequencies of geodesic motion at a given orbit, scale inversely with the mass~$M$ of the central compact object when the dimensionless forms of the neutron star angular momentum~$j$ and quadrupole moment~$q$ are held fixed. Therefore, within the framework of the above models, if the absolute differences between the squares of angular momenta are small among the sources being considered, the ratio $R=\nuU/\nuL$ between the twin kHz QPO frequencies represents a rough measure of the radial position of the QPO excitation, which is independent of the mass of the neutron star \citep[][]{tor-etal:2008:aca:stella}. The frequency ratio~$R$ also has genuine importance for the resonance~model.

\subsection{Strength of the signal}\label{strength}

Because of the expected links to the orbital motion, most discussions of neutron star twin kHz~QPOs have for a long time been concentrated mainly on the frequencies $\nuL,~\nuU$, and on their relations and evolution.

The other QPO properties, namely the \emph{quality factor}~$Q$ and \emph{fractional root-mean-squared (rms) amplitude}~$r$ have also been studied, but have not attracted such wide attention. We recall that the quality factor~$Q$ characterises the coherence time of a QPO, being defined as the QPO centroid frequency over the full-width of the peak at its half-maximum, while the {fractional root-mean-squared amplitude}~$r$ represents a measure of the signal strength, which is proportional to the square root of the peak power contribution to the PDS. In the past few years, both the quality factor and the rms amplitude have been studied systematically for several sources, and possible consequences for various QPO models have been outlined~\citep[see, e.g.,][~for further information and 
references]{Men:2006:MONNR:371,Bar-Oli-Mil:2006:MONNR:QPO-NS}.

 However, most attention has been focussed on the quality factor and the rms amplitude as separate functions of frequency. Little attention has been paid to the \emph{mutual relations} between the (correlated) QPO amplitudes.

In the present work we study the difference in strength between the twin kHz QPOs, $\Delta r= r_{\L}-r_{\U}$,
defined as the difference between the fractional rms amplitudes of the lower and upper kHz QPOs ($r_\L, r_\U$).

%
\section{Data analysis}\label{sixatolls}

Here we focus on six well-studied atoll sources (namely \mbox{4U~1608$-$52}, \mbox{4U~1636$-$53}, \mbox{4U~0614+09}, 
 \mbox{4U~1728$-$34}, \mbox{4U~1820$-$30}, and \mbox{4U~1735$-$44}) spanning a wide range of 
 frequencies.

%
\begin{figure*}[t!]
\begin{minipage}{1\hsize}
\begin{center}
\hfill
\includegraphics[width=.0465\textwidth]{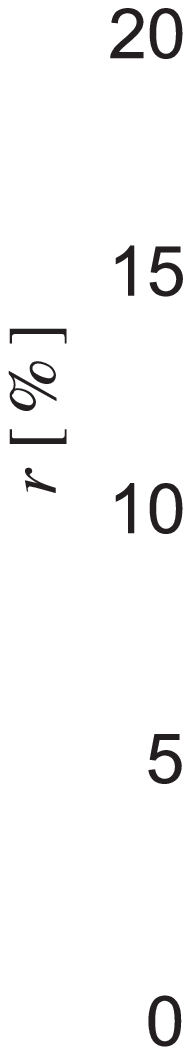}
\includegraphics[width=.315\textwidth]{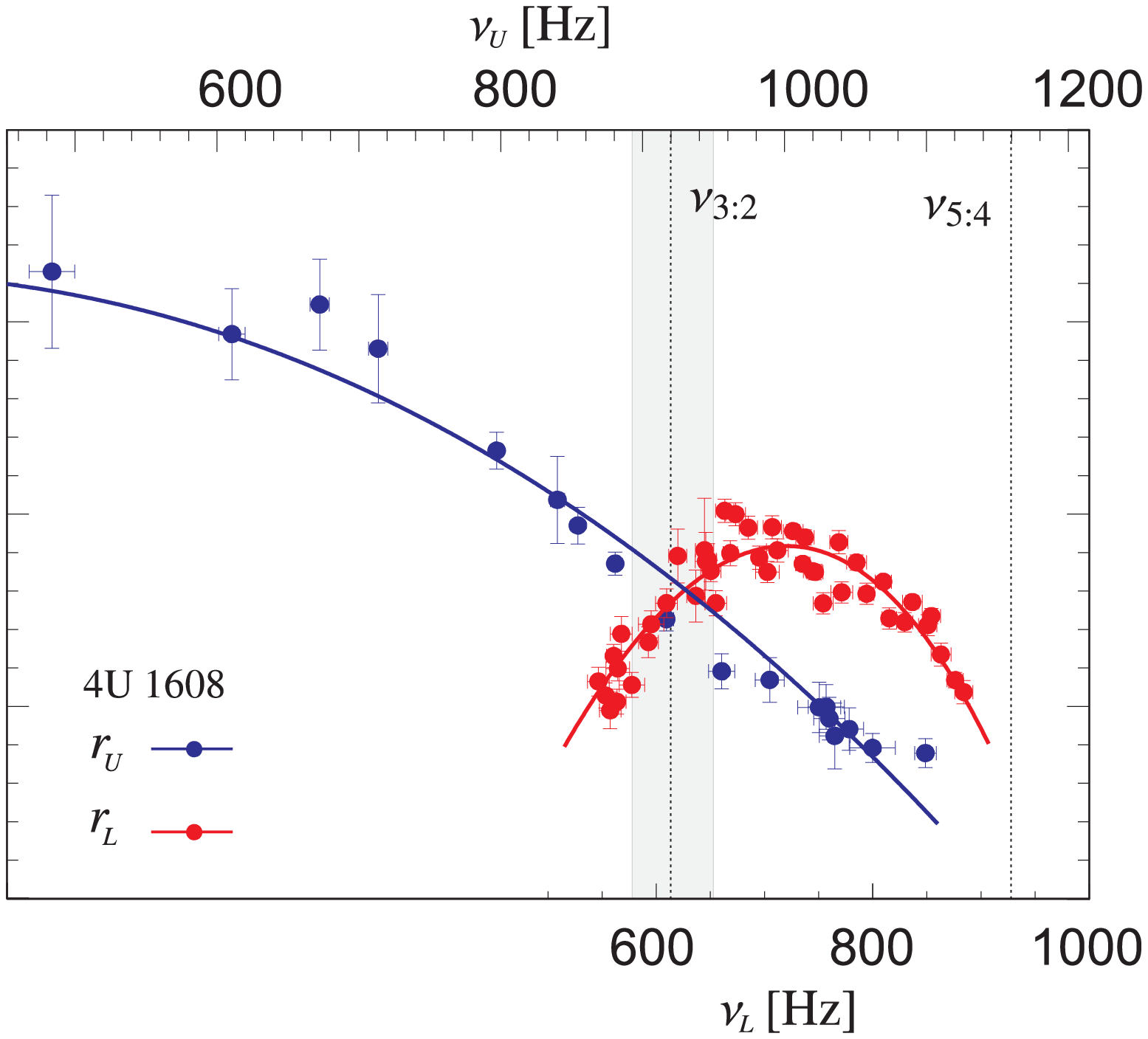}\hfill
\includegraphics[width=.315\textwidth]{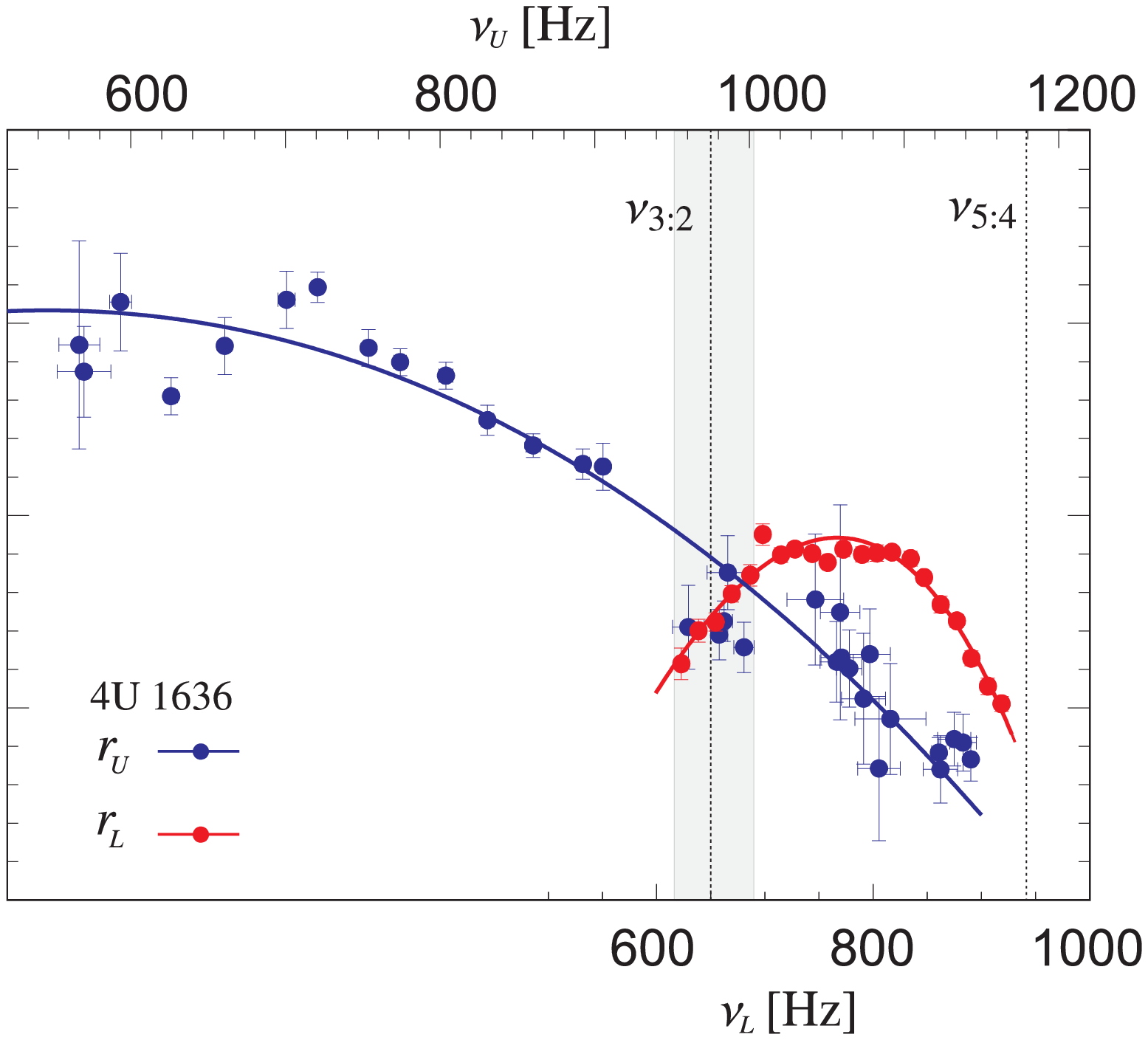}\hfill
\includegraphics[width=.315\textwidth]{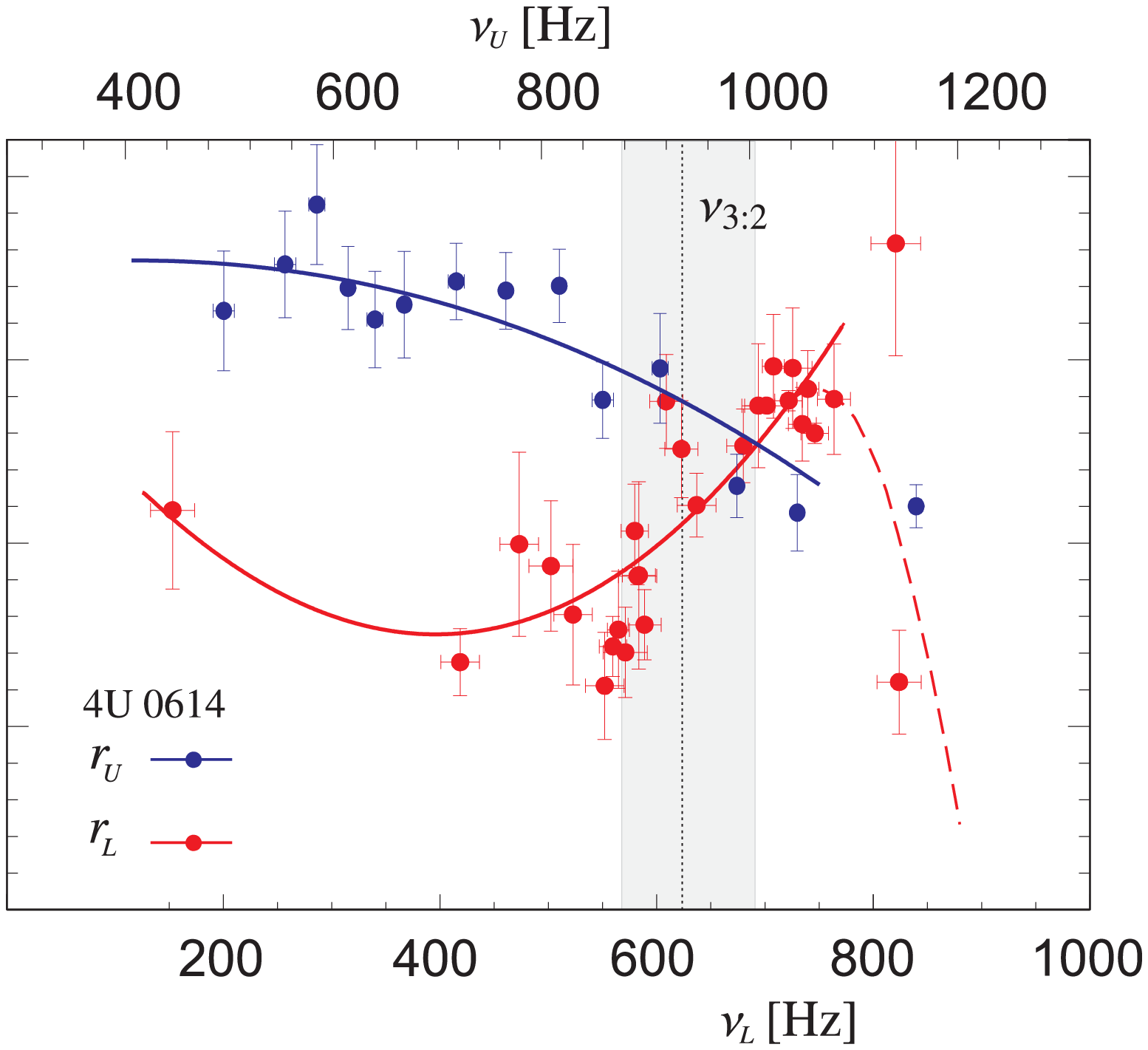}\hfill

\noindent
\hfill
\includegraphics[width=.0465\textwidth]{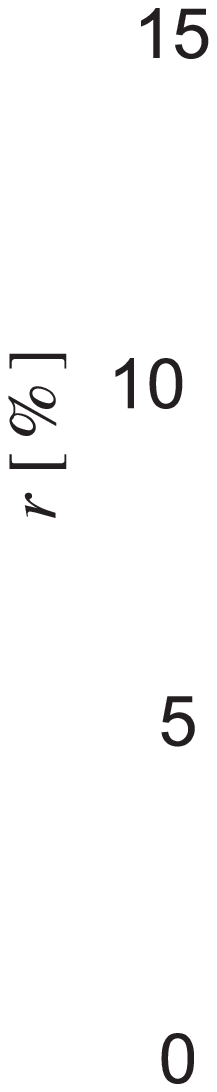}
\includegraphics[width=.315\textwidth]{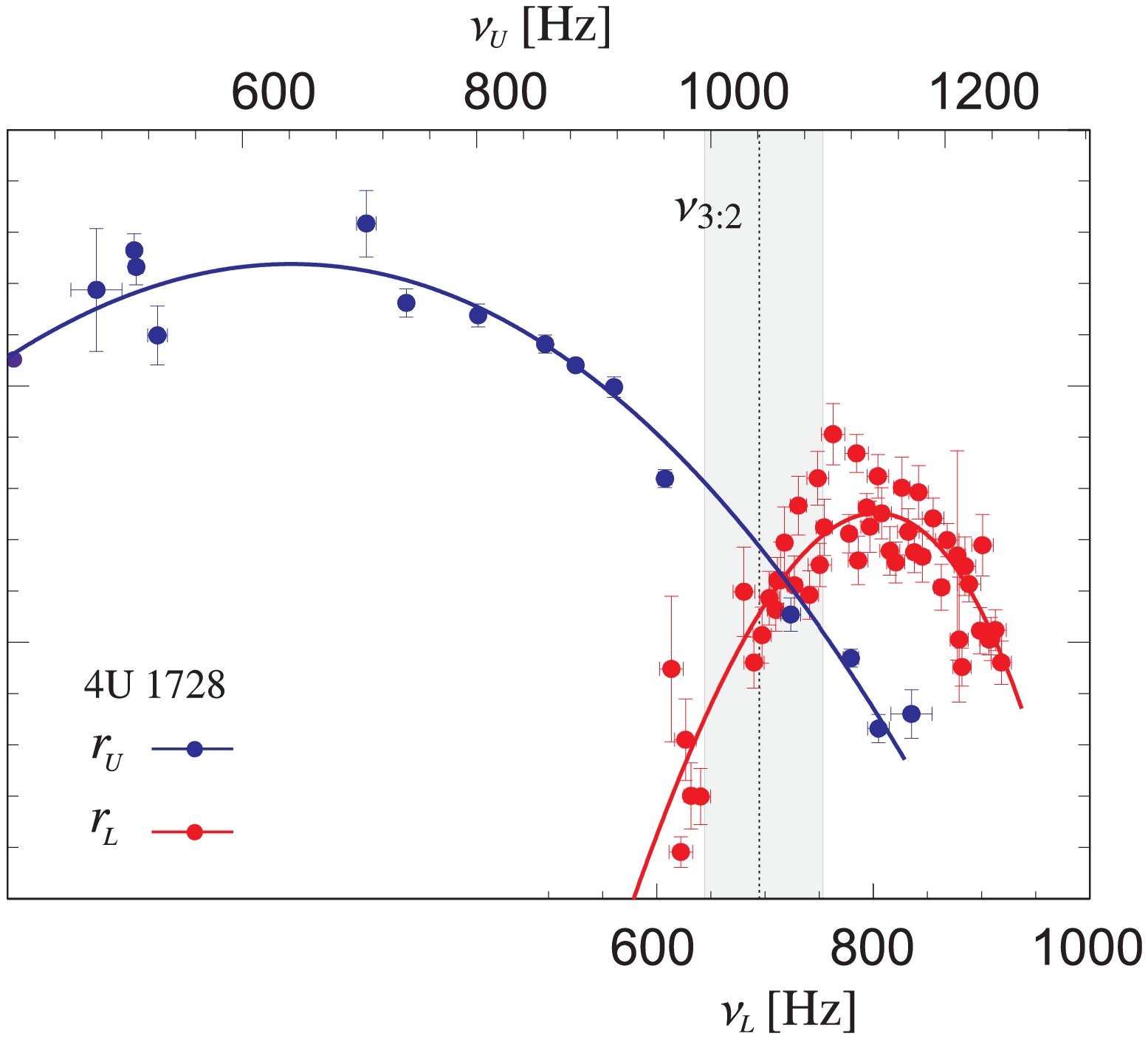}\hfill
\includegraphics[width=.315\textwidth]{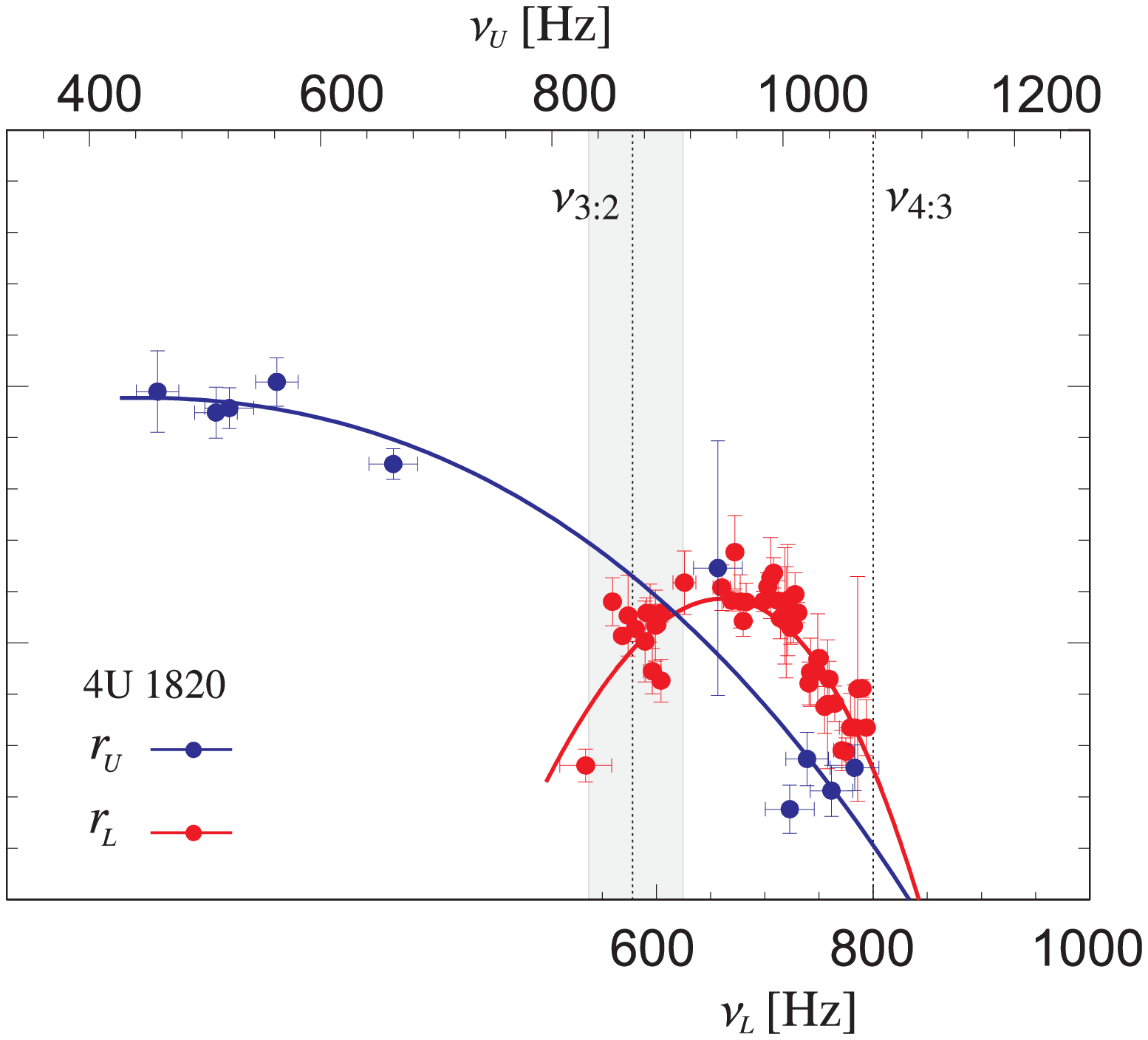}\hfill
\includegraphics[width=.315\textwidth]{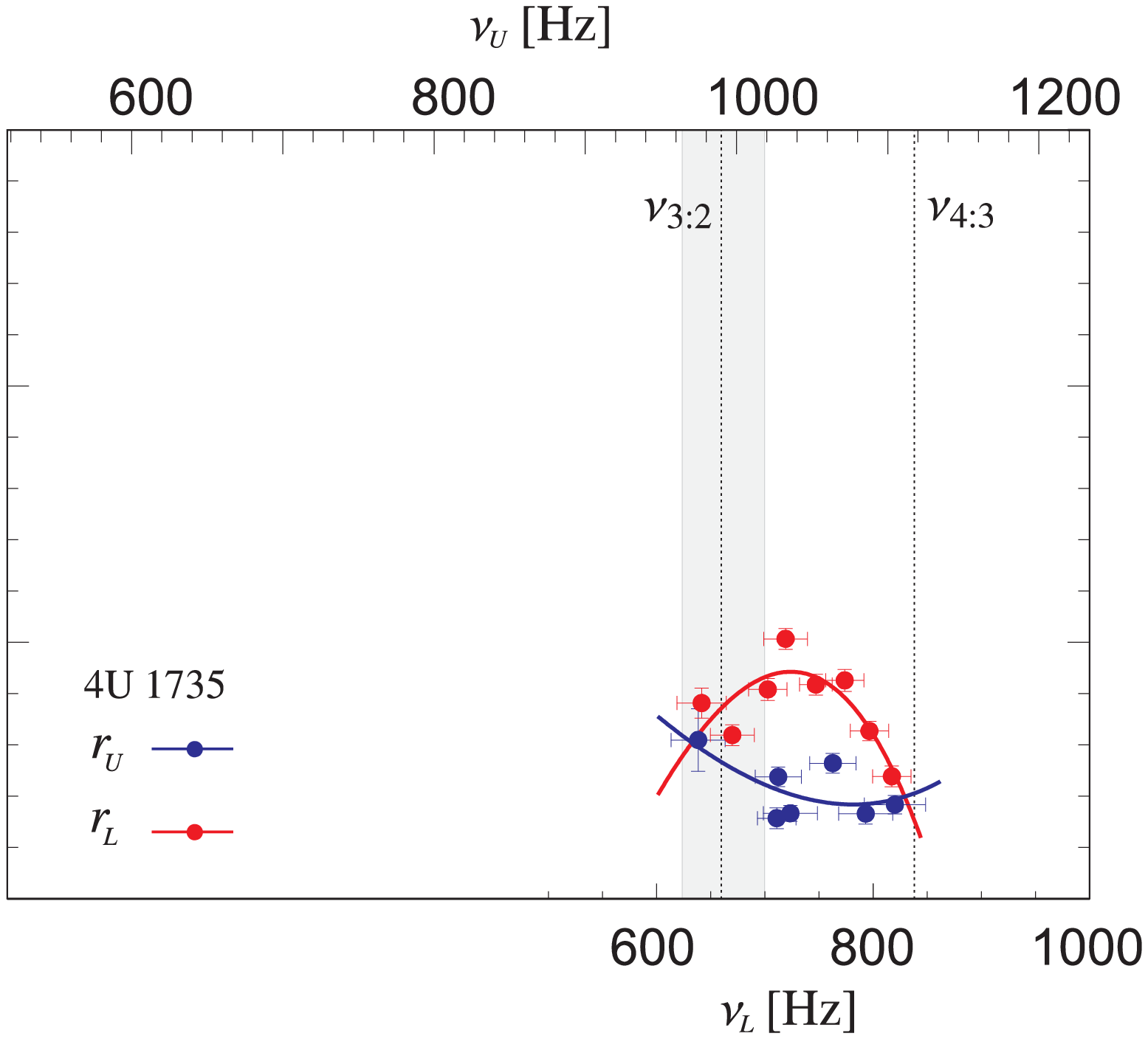}\hfill

\end{center}
\end{minipage}
\caption{\label{figure:r} Amplitudes of the lower and upper oscillations
  plotted against\ the two QPO frequencies (published data).  Above 
  $\nuL\sim\!500\,\mathrm{Hz}$~(i.e., below $R\gtrsim1.7$), the frequency 
  correlations are approximated well by the linear relations that we use. 
  The shadow indicates the frequency range corresponding to $R\!=\!1.5\pm3\%$. 
    For the lower QPO in 4U~0614+09 the high-frequency part contains two datapoints that do 
  not indicate a clear trend.}
\end{figure*}
%
%

\subsection{Published data (Set I)}
\label{section:published_data}
  
We benefit from existing studies and use published results on rms amplitude correlations from the large collection made by~\citet{Men:2006:MONNR:371}. We refer to this data as \emph{Set I}. The main references to the original papers are as follows:
  4U~1608-52~\citep{Men-Kli-For:2001:ASTRJ2:AmplQPO},
  4U~1636-53~\citep{Bar-Oli-Mil:2005:MONNR:},
  4U~0614+09~\citep{Str-etal:2002:ASTRJ2:MultiLor},
  4U~1728-34~\citep{Men-Kli-For:2001:ASTRJ2:AmplQPO},
  4U~1820-30~\citep{Bar-Oli-Mil:2006:MONNR:QPO-NS},
  4U~1735-44~\citep{Bar-Oli-Mil:2006:MONNR:QPO-NS}. Although obtained 
in different studies (using different approaches), the data are comparable 
since they are measured over the whole RXTE-PCA energy band 
(2$\fromto$60\,keV).

The data of Set I do not always correspond directly to simultaneous twin kHz QPO measurements. Moreover, the 
range where the properties of either one of the QPO frequencies is well-investigated often extends far from the region of joint coverage, and then the identification of a particular significantly-detected peak as belonging to 
the upper or lower group comes from an extensive investigation of its properties (especially its quality factor) and also its correlations with the low-frequency QPOs~\citep[see][]{Men-Kli-For:2001:ASTRJ2:AmplQPO,oli-etal:2003,Bar-Oli-Mil:2005:MONNR:,Bar-Oli-Mil:2005:AN:,Bar-Oli-Mil:2006:MONNR:QPO-NS,bel-etal:2007,Kli:2006:CompStelX-Ray:}. It is relevant in this context that both QPOs become weak broad features when their frequencies are lower than $\sim\,$500\,Hz \citep[see][]{oli-etal:2003}.

For the purposes of this paper, we interpolate the amplitude data of Set I for each source using the best fits for both amplitudes made in terms of a sum of three exponentials
\begin{equation}
 r_i=\sum_{j=1}^3\left[\exp
      \left(P_{i,j}^{(1)}+\nu_{i} P_{i,j}^{(2)}\right)+P_{i,j}^{(3)}
    \right]\,,\quad i\in\{\U,\L\}\,,
\label{equation:cermacek}    
\end{equation}
with the $P_{i,j}^{(k)}$ being free parameters.

As noticed in several studies, the frequency correlations in the six sources considered in this paper are well-fitted by linear relations~\citep{Abr-etal:2005:ASTRN:,Bel-Men-Hom:2005:ASTRA:,Bur:2006:PhD:,Zha-etal:2006:MONNR:kHzQPOFrCorr}.
\emph{We use this property to connect both amplitude-frequency relations (\ref{equation:cermacek}), with the assumption that the frequency correlations are linear in the range above $\nuL=500\,\mathrm{Hz}$, and follow the relations found in~\citet{Abr-etal:2005:ASTRN:,Abr-etal:2005:RAGtime6and7:CrossRef}.} These relations were determined from the well-investigated twin QPO data above $\nuL\!=\!500\,$Hz~, i.e., below $R\!\simeq\!1.7$. {The highest frequency ratio detections in the atoll sources being considered are around $R\sim\!2.5\fromto3$.\footnote{See \cite{Zha-etal:2006:MONNR:kHzQPOFrCorr} who study frequency correlations in a large group of sources using for the atoll sources the data from works of \cite{bel-etal:2002,Bel-Men-Hom:2005:ASTRA:}, \cite{psa-etal:1999a,psa-etal:1999b}, \cite{Kli:2000,Kli:2006:CompStelX-Ray:}, \cite{men-etal:1998,men-etal:1998b,men-etal:1998c}, \cite{men-kli:1999}, \cite{str-etal:2000,Str-etal:2002:ASTRJ2:MultiLor,str-etal:2003}, \cite{sal-etal:2003}, \cite{jon-etal:2002}, \cite{mark-etal:1999}, \cite{mig-etal:2003}.} However, the detections above $R\sim\!2$ are rare and an extension to the related range \mbox{$\nuL\!\sim\!200\fromto400$\,Hz} is subject to errors in determining the above relations.}

\subsection{Data processed in a uniform way (Set II)}

Following~\citet{Bar-Oli-Mil:2005:MONNR:,Bar-Oli-Mil:2006:MONNR:QPO-NS}, we use 
all public archival RXTE-PCA data available up to the end of 2004 (for the full 
energy band from 2$\fromto$60\,keV) for the six atoll sources that we are 
considering. Segments of the temporally continuous collection of data from a 
single pointing are processed using a shift-add procedure. For each segment, a 
set of 8-second leahy normalized PDS is produced, and an averaged PDS is then 
calculated for each set. The averaged PDS is searched for QPOs using a scanning 
technique that looks for peak excess (assumed to have a Lorentzian profile) 
above the Poisson counting noise level~\citep[see][]{Boi-etal:2000:}. To 
minimize the effects of long-term frequency drift, the QPO frequencies are 
estimated on the shortest statistically permitted timescales, applying a 
recursive search algorithm seeking over the set of 8-second PDS. If there are 
two QPO peaks present in the segment-averaged PDS, the peak with the highest 
\emph{significance}~$S$ (defined as the integral of the Lorentzian fitting of 
the peak divided by its error) is considered for further tracking. The tracked 
QPO is searched for within a window with a width not greater than $20\%$ of the QPO frequency. {The present frequency-ratio detections in the sources being considered are above $R\!=\!1.2$. Because the relative frequency change within a single observation is small, the algorithm cannot switch between the two peaks within the tracking.}
{After the tracking, the segment splits into several averaged subsegments (each of them corresponding to about 50$\fromto500$\,sec) to which the shift-add is applied.} The PDS obtained is then 
searched for the final QPOs. For later analysis, we retain the resulting 
peaks detected above the threshold $S \geq 2.5\sigma$. We refer to these data, representing a coherent set processed in a uniform way, 
as Set~II.

\section{Results}\label{adifdir}%

\begin{figure*}[t!]
\begin{minipage}{1\hsize}
\begin{center}
\includegraphics[width=1\textwidth]{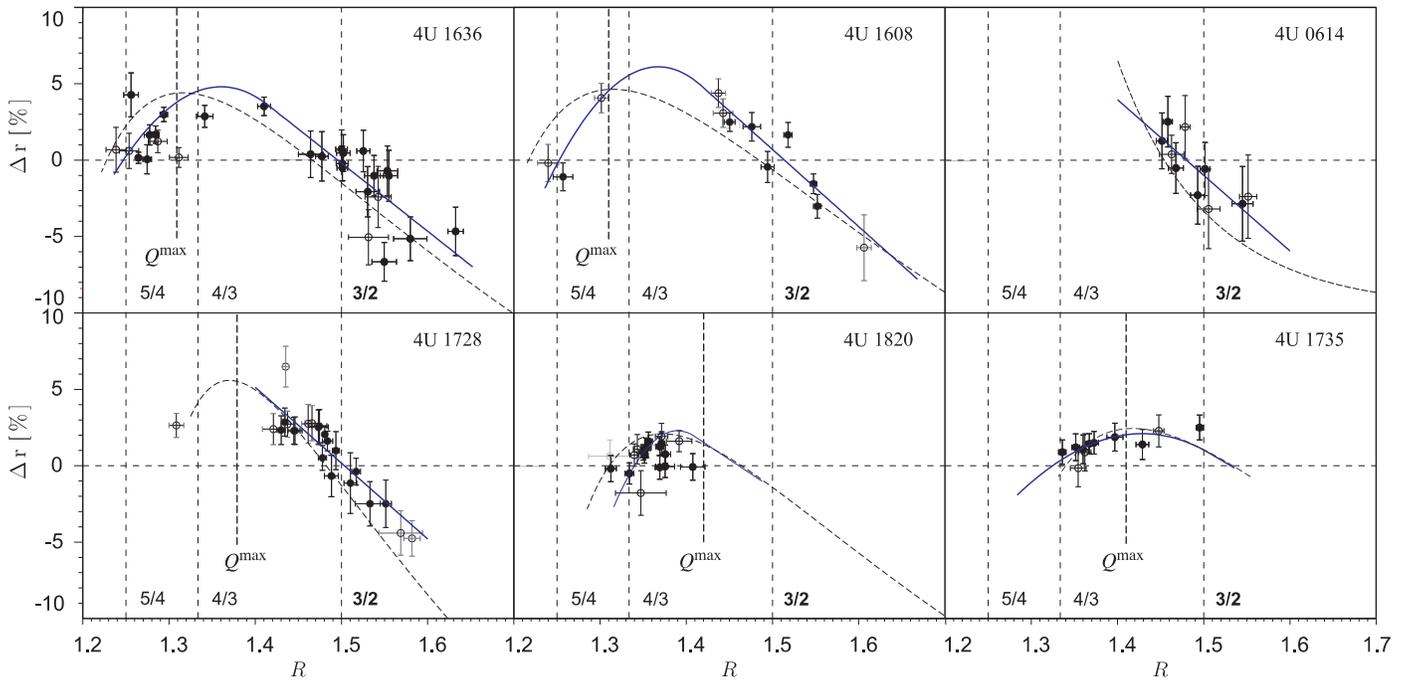}
\end{center}
\end{minipage}
\caption{\label{figure:rc2} The dependence of the amplitude difference on the 
frequency ratio, with detailed view shown of the range $R\!=\!1.2\fromto1.7$. 
The dashed black curves correspond to Set I (with exponential interpolation), 
while the individual points correspond to Set II. When there are asymmetric 
errors, we always use the larger one. The open points are above the $2.5\sigma$ 
threshold; the filled points have significance greater than $3\sigma$. The 
solid blue curves use linear interpolation in the decreasing part of curve 
joined to an exponential in the increasing part (see text). The vertical lines 
labelled with {$Q^{\mathrm{max}}$} denote the frequency ratio corresponding to 
the maximum of the lower QPO quality factor (see Section 
\ref{section:discussion} for a comment on this). }
\end{figure*}
%
%

It can be seen from Fig.~\ref{figure:r} that the rms amplitudes of the two 
peaks are comparable over a frequency range that is rather narrow with respect 
to the total frequency range spanned by the six sources. Moreover, for all of 
the six sources, the rms amplitudes are comparable in a range around the 
frequency where the $3:2$ ratio occurs. We mark this frequency and those for 
other relevant ratios in Fig.~\ref{figure:r}.

Consideration of individual frequency correlations implies that lower and upper QPO datapoints overlaps in Fig.~\ref{figure:r} do fall mostly within the range $R\!\sim\!1.2\fromto1.7$. In Fig.~\ref{figure:rc2} we show the behaviour of the amplitude difference $\Delta r$ in this range, as inferred from the interpolated curves in Fig.~\ref{figure:r}. In the same figure we also plot values of $\Delta r$ from Set II. We detected only three significant datapoints outside this range in Set II. {Each of these three datapoints falls into the range $R\in[2.2,2.7]$ and corresponds to negative values of $\Delta r$}.

\subsection{Roots of $\Delta r$}
\label{section:roots}

The values of $\Delta r$ at the intersection with the $R$ axis are shown in Table \ref{table:1}. We refer to the intersections as \qm{roots $R_1$, $R_2$}, corresponding to the higher and lower frequency 
ratios.

In the four sources (4U~1608-52, 4U~1636-53, 4U~0614+09, 4U~1728-34) where we detected enough datapoints in Set~II between $R=1.4\fromto1.6$, the values of $\Delta r$ are well-approximated by linear functions of $R$ in that range. We then fit these datapoints by straight lines (see Table \ref{table:1} for the corresponding $\chi^2$). For 4U~1636-53 and 4U~1608-52, which display in the Set II data a clear increase in 
amplitude difference between $R=1.2\fromto1.4$, we join the straight-line fit 
to an increasing exponential fit. All of the resulting curves for Set II are 
included in Figure \ref{figure:rc2}.

For the four sources 4U~1608-52, 4U~1636-53, 4U~0614+09, and 4U~1728-34, the roots $R_1$ for the Set II 
data given in Table \ref{table:1} are determined from the fits described above. 
In principle, the fits can enable us also to determine the roots $R_2$ from the 
Set II data for 4U~1636-53 and 4U~1608-52. These are close to $R\sim1.25$ but are difficult 
to quantify precisely. Clearly, as can be seen from Fig. \ref{figure:rc2}, 
more investigation is needed in order to make more precise statements about 
these two roots. In 4U~1608-52 there are only three relevant datapoints, while in 
4U~1636-53 the datapoints give a rather high $\chi^2$ for the increasing part of 
$\Delta r$ curve.

For 4U~1820-30 and 4U~1735-44, there is a lack of twin kHz QPO detections around 
$R\!\sim\!1.5$ from Set II, and we were not able to determine their roots 
$R_1$, although the Set I data indicate that they should exist. For the 
increasing part of the $\Delta r$ curve, we use the same fitting procedure as 
for 4U~1608-52 and 4U~1636-53, but we only \emph{expect} the existence of the linear 
decaying part between $R\!=\!1.4\fromto1.6$, which we determined from the Set I 
data. In this way, we find the two roots $R_2$ close to $1.33$. For 4U~1820-30, the 
datapoints close to $R_2$ are scattered around the $R$-axis, giving a rather 
poor $\chi^2$ for all of the fits, and the value of $R_2$ varies with small 
changes of $\chi^2$. For 4U~1735-44, the location of $R_2$ seems to be 
well-determined.

We note that for 4U~1820-30 and 4U~1735-44 ($R_2\sim1.33$), as well as for 4U~1608-52 and 4U~1636-53 ($R_2\sim1.25$), the termination of the available data nearly seems to coincide 
with the position of $R_2$, which makes the exact determination of $R_2$ 
difficult.

%
\definecolor{slightlyGray}{gray}{0.85}
\newcommand {\graybox} [1] {\raisebox{0pt}[0pt]{\fcolorbox{black}{slightlyGray}{#1}}}
\newcommand {\grayarrea} [1] {#1}
\begin{table*}[t!]
\caption{\label{table:1}The roots $R_{1}$, $R_{2}$ of the function~$\Delta r(R)$.}
\begin{center}
\begin{tabular}{lccccccc}
\hline
Source~~~~~&\multicolumn{4}{c}{Set~I}&\multicolumn{3}{c}{Set~II}\\
&$\chi^2_\L/d.o.f$&$\chi^2_\U/d.o.f$&$R_1(\Delta r = 0)^\dagger$&$R_{2}(\Delta r = 0)$
&{$\chi^{2}/d.o.f$}&$R_{1}(\Delta r = 0)$
&$R_{2}(\Delta r = 0)$\\
\hline\vspace{-2ex}\\
4U~1608-52 &17/38&15/9& $1.48\pm0.01$ & -- &{13/8}$\,^*$& \grayarrea{$1.51\pm0.01$} & $1.25\pm0.01$  \\
4U~1636-53 &1.7/11&40/23& $1.49\pm0.01$ & -- &{18/15}$\,^*$& \grayarrea{$1.49\pm0.01$} & $1.25\pm0.01$  \\
4U~0614+09 &59/15&16/4& $1.45\pm0.01$ & -- &{3.8/10}$\,^*$& \grayarrea{$1.48\pm0.01$} & --  \\
4U~1728-34 &40/36&6.3/7& $1.48\pm0.01$ & -- &{21/23}$\,^*$&\grayarrea{$1.50\pm0.01$} & --  \\
4U~1820-30 &23/35&15/7& $1.46\pm0.01$ & $1.31\pm0.02$ &32/15& -- & $1.34\pm0.02$  \\
4U~1735-44 &2.6/4&2.5/2& $1.53\pm0.02$ & $1.34\pm0.01$ &4.1/8& -- & $1.33\pm0.01$\\
\hline
\multicolumn{8}{l}{$^\dagger$the errors shown correspond to a unit variation of~$\chi^{2}$; for asymmetric errors, we use the larger one.}\\
\multicolumn{8}{l}{$^*$\,corresponds to the linear part of curve between $R\!\!=1.4\fromto1.6$.{\Large$\phantom{A}$}}
\end{tabular}
\end{center}
\end{table*}
%
%

%
\section{Discussion and conclusions}\label{section:discussion}

In five of the six atoll sources that we have considered (\mbox{4U~1728$-$34}, 
\mbox{4U~1608$-$52}, \mbox{4U~1636$-$53}, \mbox{4U~0614$+$09}, and \mbox{4U~1820$-$30}), the upper kHz QPO dominates the lower one for high values of the twin kHz QPO ratio ($R\!>\!1.5$)
corresponding to low QPO frequencies. The issue of the upper QPO dominance at low QPO frequencies in \mbox{4U~1735$-$44} remains an open question due to the lack of the relevant datapoints in both Sets I and II. The four sources \mbox{4U~1728$-$34}, \mbox{4U~1608$-$52}, \mbox{4U~1636$-$53}, and \mbox{4U~0614$+$09} clearly exhibit the $R_1$ root close to $R\!=\!1.5$. Because of the lack of datapoints in Set II, however, more investigation is needed to fully confirm a similar statement about $R_1$ for the sources 4U~1820-30 and 4U~1735-44. For ratios decreasing from the value $R\simeq1.5$, the difference between the QPO amplitudes reaches its maximum within a narrow interval of $R\!\in\![1.3,~1.45]$.\footnote{{\cite{Bar-Oli-Mil:2005:MONNR:,Bar-Oli-Mil:2005:AN:,Bar-Oli-Mil:2006:MONNR:QPO-NS} determined the frequencies corresponding to maxima of the lower QPO quality factor in the discussed sources. Interestingly, the relevant ratios are close or coincide with those related to the maxima of $\Delta r$  (see Fig.~\ref{figure:rc2}).}} After this, as the ratio decreases further (but not for 4U~0614+09 where the available data terminates), the lower QPO starts to lose its dominance again. Notably, all of the determined values of the root $R_1$ lie inside the $\sim\!3\%$ interval around $R\!=\!1.5$, which corresponds to less than $10\%$ of the range $R\in(1.2,\,2.5)$. For the four sources (4U~1608-52, 4U~1636-53, 4U~1820-30, 4U~1735-44), and marginally also for 4U~1728-34, there is evidence for of another root (close to~$R= 1.33$ or~$1.25$) where the available data terminate. {Most of datapoints in Fig. \ref{figure:rc2} appear close to the roots $R_1$ or\,/\,and $R_2$. We therefore suggest that there could be a link between the previously reported ratio clustering and the existence of $R_{1,2}$, which requires further investigation.}
 

Within the framework of the QPO resonance models, the behaviour of the 
amplitude difference may indicate an energy interchange between 
the lower and upper QPO modes, typical of non-linear 
resonances~\citep[e.g.,][]{Hor-Kar:2006:ASTRA:TOQPOIntRes}. In this context, it 
is also rather noticeable that the roots $R_{1,\,2}$ are \emph{close} to the 
values $3/2$ and $4/3$ or $5/4$ (which we denoted by dashed vertical lines in 
the above figures). Nevertheless, in order to take this indication seriously, 
it is necessary to have a model that gives a detailed explanation of the effect.

Within the scope of the other orbital QPO models, there should also be an 
explanation for why the roots $R_{1,\,2}$ appear. For instance, in the case of 
the relativistic precession model, one can solve equations relating the upper 
and lower QPO with respect to the radial coordinate $\rho$, for the particular 
form of the expressions for the orbital frequencies in Schwarzschild geometry, 
obtaining the 
relation~$\rho\!=\!{6MR^{2}}/{(2R-1)}$, see \cite{tor-etal:2008:aca:stella}. According 
to this formula, the~$R\!=\!1.5$ frequency ratio would correspond to an orbital 
radius~$\rho\!=\!6.75\,M$, which is about~14~kilometres for a neutron star with the 
``canonical'' mass~$M=1.4\,M_{\sun}$. The~$3\,\%$ scatter around $R\!=\!1.5$ 
would be projected onto an interval of only about~$\pm1.5\,\%$ in~$\rho$, i.e., 
onto~$\sim\pm0.1\,M\doteq\pm200\,\mathrm{m}$. It is then a question not only of why the observed amplitudes corresponding to oscillations at this radius should be equal, but also why they are equal for several sources at the same radius with a scatter of less than~$250\,\mathrm{m}$.

\acknowledgements
I would like to thank Didier Barret for all his personal, scientific, and technical help, for his patience and for providing the data and analysis software on which this paper builds. Then I thank Marek Abramowicz, Michal Bursa, Ji\v{r}\'{\i} Hor\'{a}k, W{\l}odek Klu\'zniak, Zden\v{e}k Stuchl\'{i}k, and Eva \v{S}r\'{a}mkov\'{a} for discussions and help with the paper. I am especially grateful to Pavel Bakala and Petr \v{C}erm\'{a}k for many debates and technical advice, and John Miller for the time spent on a detailed discussion of the paper and its wording. I also express sincere thanks to the anonymous referee for his/her critical comments and suggestions that greatly improved the quality of the paper. This work was supported by the Czech grant MSM 4781305903.



\end{document}